\documentclass[twocolumn,aps,prb,showpacs,amsmath,amssymb, floatfix]{revtex4}
\usepackage[dvips]{graphicx}
\usepackage{dcolumn}
\usepackage{bm}
\usepackage{color}

\begin{document}
\title{Electrical Detection of Spin Excitations}
\author{Sayeef Salahuddin and Supriyo Datta}
\affiliation{School of Electrical and Computer Engineering, Purdue University, West Lafayette, IN-47907, USA.}
\date{\today}

\begin{abstract}
The objective of this paper is to draw attention to a possible new approach for measuring the spin excitation spectrum of a spin array by placing it in intimate contact with the channel of a spin-valve device in an anti-parallel (AP) configuration. We show, using realistic device parameters, that the second derivative of the current $\partial^2I/\partial V^2$ should exhibit peaks whose location corresponds to the energy of excitation and whose height is its strength. The effect is maximum for ideal half metallic contacts in AP configuration but we show that even with contacts that are less than ideal, the spin signals can be extracted using a de-embedding scheme. Compared to existing techniques our proposal has three attractive features: (a) high sensitivity down to single spins, (b) high spatial resolution limited only by the wavefunctions of the individual spins and (c) a relatively common device structure to which a lot of current research is devoted  
\end{abstract}

\maketitle
In recent years there has been increased interest in the dynamics of spin arrays, partly motivated by the possibility of using such arrays for a wide range of applications, such as quantum computing\cite{kanenature:ref, burkard:ref, vrijen:ref,zutic:ref}, data storage \cite{gurney:ref}, biological sensing \cite{borkowski_science:ref}, DNA tagging and targeted drug delivery \cite{hoffman:ref}. Except for data storage, these applications require the ability to detect the spin spectra of only a few spins. This represents a major challenge\cite{rugar:ref, jiangpaper:ref} and there has been a number of new approaches which have attempted to improve on the standard techniques of spin detection which rely on detecting electronic or nuclear spin resonance \cite{opticsdetection1:ref,opticsdetection2:ref,magneticdetection:ref,scanningtunneling1:ref,loss:ref,tsoi:ref}. The current state of the art is magnetic resonance force microscopy in which single spin resonance detection \cite{rugar:ref} has been recently reported. An alternative approach is to couple the resonance to the current flow in an adjoining conductor. Engel and Loss \cite{loss:ref} proposed a method to detect the single spin decoherence inside a quantum dot by looking at the charge current. More recently, Xiao et. al \cite{jiangpaper:ref} reported detection of single spin resonance by looking at the current that flows through a conventional Si MOSFET. 

However, a common problem of any type of resonance detection is the spatial resolution \footnote{A single shot measurement of a single spin, where ESR is not needed, was reported recently \cite{kowenhoven:ref}. However, this method involves a very slow measurement method (normally a few ms for each measurement). }. With the standard techniques, it is very difficult to achieve a resolution below 1$\mu$m \cite{rugar:ref}. Even with magnetic force resonance microscopy, which was originally proposed to improve the spatial resolution, the state of the art is 25 nm. The lack of spatial resolution a well-known matter of concern in detecting very small magnetic particles using MRI \cite{chung:ref}. For quantum computing algorithms too, one often needs to be able to readout a specific portion of the spin array with high spatial resolution. 

\begin{figure}[b]
	\centering
	\includegraphics[width=6cm]{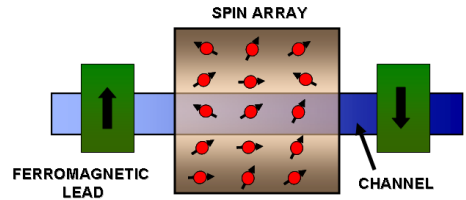}
  \caption{A schematic showing the top view of the proposed device. The channel is connected to ferromagnetic/halfmetallic leads arranged in an anti-parallel configuration. A spin array, is placed in close proximity of the channel and affects the conduction by spin flip scattering through exchange coupling with channel electrons.}
	\label{FIG1}
\end{figure}

In this paper, we show that a lateral conductor having spin-polarized contacts in AP configuration (see Fig. \ref{FIG1}) allows one to detect a very small (even one) number of spins with a spatial resolution limited only by the extent of the wave function of the individual spins. Ever since the earliest proposal for a ``spin-transistor"\cite{datta-das:ref}, devices with ferromagnetic contacts have been the focus of much research effort. However, most of the work is devoted towards making parallel contacts. We show that such parallel configurations are unable to detect the spin spectra, which should show up if we use the AP configuration instead. In this configuration, the second derivative of the current with respect to voltage ($\partial^2I/\partial V^2$) provides a direct measurement of the spin excitation spectrum. However, if the contacts do not have perfect injection efficiency, the spin signatures may get mixed up with contributions from other effects e.g. electron-phonon interaction. For this case, we propose a `de-embedding' technique that can recover the spin information.

Fig.\ref{FIG1} shows the proposed device structure which could be a spin-valve device \cite{gurney:ref} or a commercial FET with ferromagnetic( ideally half-metallic) contacts. The spin array could, for example, be the paramagnetic traps at the Si-SiO$_2$ interface of an FET \cite{jiangpaper:ref}, ferromagnetic atoms trapped inside a host molecule attached to the channel \cite{kondo:ref}, or a ferromagnetic gate with good exchange coupling with the channel electrons. 

\emph{Ideal Contact.-}  First we assume ideal half metallic contacts such that the source injects (and detects) only `+z'-polarized spins, while the drain injects (and detects) only `-z'-polarized spins. The current can flow only if electrons can flip their spins by interacting with a spin array.The point is that \emph{in this idealized structure any current that flows is a measure of the spin-flip scattering in the channel region and as such provides a measure of the excitations of the spin array}. In our model, we assume the spin-scattering is caused by an exchange interaction between a channel electron and individual spins of the array. The form of the exchange interaction is taken to be $\sum_{R_j}J(\bar{r}-\bar{R_j})\vec{\sigma}\cdot\vec{S}_j$ where, $r$ and $R_j$ are the spatial coordinates and $\sigma$ and $S_j$ are the spin operators for the channel electron and $j$-th spin in the array. $J(\bar{r}-\bar{R_j})$ is the interaction constant between the channel electron and the j-the spin in the array. The current flow in the device is proportional to the difference between the rates at which the channel electrons are scattered from `-z'$\rightarrow$`+z'(denoted by $F_1$) and `+z'$\rightarrow$`-z'(denoted by $F_2$):
\begin{equation}\label{scattering_current}
I_s=e\frac{2\pi}{\hbar}\left(F_1-F_2\right)
\end{equation}
where $F_1$ and $F_2$ are given by
\begin{subequations}
\begin{equation}\label{F1}
\begin{split}
F_{1}=&\sum_{\alpha,\beta,m\uparrow,n\downarrow, R_j}|<m\uparrow,\beta|J(\bar{r}-\bar{R_j})\vec{\sigma}\cdot\vec{S_j}|n\downarrow,\alpha>|^2 \\
&\delta(E_\beta+E_{m\uparrow}-E_\alpha-E_{n\downarrow})P_\alpha f_{n\downarrow}(1-f_{m\uparrow})
\end{split}
\end{equation}
\begin{equation}\label{F2}
\begin{split}
F_2=&\sum_{\alpha,\beta,m\uparrow,n\downarrow,R_j}|<n\downarrow,\alpha|J(\bar{r}-\bar{R_j})\vec{\sigma}\cdot\vec{S_j}|m\uparrow,\beta>|^2 \\
&\delta(E_\alpha+E_{n\downarrow}-E_\beta-E_{m\uparrow})P_\beta f_{m\uparrow}(1-f_{n\downarrow})
\end{split}
\end{equation}
\end{subequations}

\begin{figure}[b]
	\centering
	\includegraphics[width=8cm]{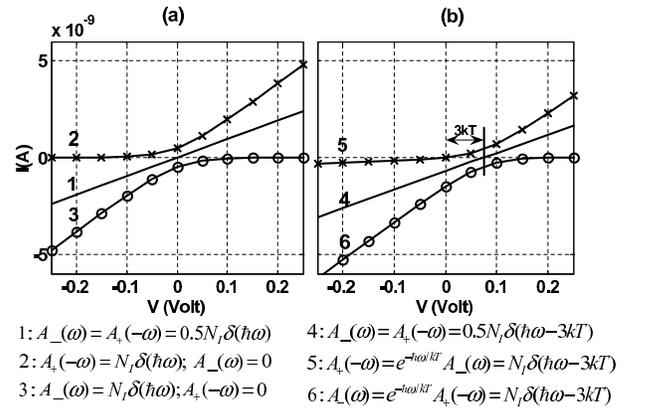}
		\caption{I-V characteristics for different spin excitations according to Eq.(\ref{scattering_current}-\ref{F2_final}) for \emph{ideal contact}.
		The simulation is done using the following parameters: $D=m^{*}LW/(2\pi\hbar^2)$, where $L=100$nm and $W=300$ nm, while $m^{*}=0.2m_0$, $m_0$ being the free electron mass. $N_I=\tilde{N}_ILW$ where $\tilde{N}_I\approx10^{11}/$cm$^2$, $T=300$K and $J_0=0.02\mu$eV which corresponds to an exchange frequency\cite{kanenature:ref} of $30$ MHz.}
	\label{FIG2}
\end{figure}

Here, $\alpha$ or $\beta$ is an eigen state of the spin array with occupation probability $P_\alpha$ such that $\sum_{\alpha}P_\alpha=1$; $m\uparrow$ and $n\downarrow$ describe the `$+$z' and `$-$z' eigenstates of the electron system with occupation functions $f_{m\uparrow}$ and $f_{n\downarrow}$ respectively. Substituting $\vec{\sigma}.\vec{S_j}=\sigma_zS_{z,j}+1/2(\sigma_+S_{-,j}+\sigma_-S_{+,j})$ and replacing sums over $m,n$ with integrals over the density of states $D(E)$, we obtain
\begin{subequations}\label{finals}
\begin{equation}\label{F1_final}
\begin{split}
F_1=\int dE& \bigg[{J_0}^2D_{\downarrow}(E) \int d(\hbar\omega)D_{\uparrow}(E-\hbar\omega)\\
&f_{\downarrow}(E)\left\{1-f_{\uparrow}(E-\hbar\omega)\right\}A_-(\omega)\bigg]
\end{split}
\end{equation}
\begin{equation}\label{F2_final}
\begin{split}
F_2=\int dE& \bigg[{J_0}^2D_{\downarrow}(E) \int d(\hbar\omega)D_{\uparrow}(E-\hbar\omega) \\ 
&f_{\uparrow}(E-\hbar\omega)\left\{1-f_{\downarrow}(E)\right\}A_+(-\omega)\bigg]
\end{split}
\end{equation}
%
\end{subequations}
where we have defined:
\begin{subequations}\label{As}
\begin{equation}\label{A_1}
A_-(\omega)=\sum_{R_j}F(R_j)\sum_{\alpha,\beta}P_\alpha|<\beta|S_{-,j}|\alpha>|^2\delta(E_\beta-E_\alpha-\hbar\omega)
\end{equation}
\begin{equation}\label{A_2}
A_+(-\omega)=\sum_{R_j}F(R_j)\sum_{\alpha,\beta}P_\beta|<\alpha|S_{+,j}|\beta>|^2\delta(E_\beta-E_\alpha-\hbar\omega)
\end{equation}
\end{subequations}
Here, $1/4|\int\psi^{*}_m(\bar{r})J(\bar{r}-\bar{R_j})\psi_n(\bar{r})d\bar{r}|^2\equiv J^2_0F(\bar{R_j})$, $J_0$ being a a constant and $F(R_j)$ representing the functional variation of interaction with $R_j$. Strictly speaking, $F(R_j)$ can depend on $m$,$n$; but we have ignored this in the following treatment. Clearly $A_-(\omega)$ and $A_+(-\omega)$ describe the spin excitation spectra of the spin array. Fig. \ref{FIG2}(a),(b) show the current vs. voltage  calculated from Eq.(\ref{scattering_current}-\ref{finals}) for an array of $N_I$ spins having excitation spectra $A_\mp(\pm\omega)$ as indicated in the figure. Note that the $A_\mp(\pm\omega)$ in case $1$ corresponds to an unpolarized (non-interacting) array, while those in cases $2$ and $3$ correspond to arrays polarized in `-z' and `z' directions respectively. Unlike case $1$, which leads to a linear I-V, cases $2$ and $3$ show rectification because a specific polarization allows current to flow by spin flip scattering only for injection from one contact but not the other. The curves in Fig.(\ref{FIG2}(a)) assume $E_\alpha=E_\beta$, whereas for the curves in Fig.(\ref{FIG2}(b)), $E_\beta-E_\alpha=\hbar\omega=3kT$, resulting in a simple shift in the I-V characteristics along the voltage axis by $3kT$.

%
    
For the plots in Fig. 2, we have assumed that $A_\pm(\mp\omega)$ remains unchanged despite current flow. This will be true if the relaxation processes (say denoted by $\gamma_I$ at $\omega$) are strong enough to keep $A_\pm(\mp\omega)$ close to their equilibrium values $A_\pm^0(\mp\omega)$ despite the spin-flip scattering process induced by the current. By writing the rate equations for $A_-(\omega)$ and $A_+(-\omega)$ and noting that $A_-(\omega)+A_+(-\omega)$ is a conserved quantity one can show that\cite{unpublished:ref}
\begin{equation}\label{steadystate}
\begin{split}
A_-(\omega)-A_+(-\omega)&=\frac{(\gamma_+-\gamma_-)}{\gamma_++\gamma_-+\gamma_I}\left\{A_-(\omega)+A_+(-\omega)\right\}\\
&+\frac{\gamma_I}{\gamma_++\gamma_-+\gamma_I}\left\{A_-^0(\omega)-A_+^0(-\omega)\right\}.
\end{split}
\end{equation}
Here, $\gamma_-$ and $\gamma_+$ are found respectively from Eqs.(\ref{F1_final}) and (\ref{F2_final}) by excluding (i)the integral over $d(\hbar\omega)$ and (ii) $A_-(\omega)$ (for $\gamma_-$) and $A_+(-\omega)$ (for $\gamma_+$). From Eq.(\ref{steadystate}), $A\pm(\mp\omega)$ will remain unaffected by the current if $\gamma_I>>\gamma_++\gamma_-$. For the parameters used in Fig.(\ref{FIG2})(see caption), this requires $\gamma_I$ to exceed $\left(\hbar/10\text{ ns}\right)\left(T/300\text{K}\right)$ which is quite likely in practical systems. Note that, in this case, the currents shown in Fig.(\ref{FIG2}) are steady state currents. But if the condition $\gamma_I>>\gamma_++\gamma_-$ is not fulfilled, the currents should be viewed as the transient currents that flow initially at t=0. Indeed, this effect can then be used to write information into the spin array \cite{datta-apl:ref,datta-italian:ref}.

The integrands of $F_1$ and $F_2$ in Eqs.(\ref{finals}) cancel each other at all energies if 
\begin{equation}
\frac{A_-(\omega)}{A_+(-\omega)}=\left(\frac{{f_{\uparrow}}^{-}}{1-{f_{\uparrow}}^{-}}\right)\left(\frac{1-{f_{\downarrow}}}{{f_{\downarrow}}}\right)=e^{(eV+\hbar\omega)/kT}
\end{equation}
where $f_{\uparrow}^{-}\equiv f_{\uparrow}(E-\hbar\omega)$. This means that the current for zero bias is zero (see Eq.(\ref{scattering_current})) if the spins are in equilibrium with the same temperature so that $A_-(\omega)/A_+(-\omega)=\text{exp}(\hbar\omega/kT)$. But if external forces  (say, for example, a resonant r.f. field), drive the spins out of equilibrium, then the zero-bias current need not be zero and under the right conditions it may be possible to extract energy from the spins and deliver it to an electrical load. For cases 2,3,4 and 6 the spin array is not in equilibrium with the thermal bath, giving rise to a non-zero current at zero bias. 

%
%
%
%
%
%
%
%

\emph{Non-Ideal Contact.-} Real systems are likely to have limited spin injection efficiency, resulting in a ballistic current in addition to the scattering current. This calls for a generalization of the expression for current (Eq.(\ref{scattering_current})). If we define the contact couplings for `$\pm$z' spins by $\Gamma_{L/R;\uparrow/\downarrow}$, it can be shown using the Non-Equilibrium Green's Function (NEGF)method that the terminal current is given by
\begin{equation}\label{terminal_prelim}
I=\frac{e}{\hbar}\int dE\bigg[D_{\uparrow}\Gamma_{L\uparrow}\left(f_L-f_{\uparrow}\right)+D_{\downarrow}\Gamma_{L\downarrow}\left(f_L-f_{\downarrow}\right)\bigg]
\end{equation}
where $f_{\uparrow}$ and $f_{\downarrow}$ have to be determined self-consistently from a combination of rate equations describing in and outflows and current conservation within the device \footnote{See Eq.(4) in \cite{datta-apl:ref}. The rate equations and resulting self-consistency were discussed for the special case where `$\pm$z' states are degenerate. The subscripts `u' and `d' in  correspond to our $\uparrow$ and $\downarrow$ respectively.}.
From Eq.(\ref{terminal_prelim}), it can be shown that 
\begin{equation}\label{terminal_current}
I=I_B+I_{s}^{'}, 
\end{equation}
where
\begin{equation}\label{terminal_first_term}
I_B=\frac{e}{\hbar}\int dE \left[\frac{\Gamma_{L\uparrow}\Gamma_{R\uparrow}D_{\uparrow}}{\Gamma_{L\uparrow}+\Gamma_{R\uparrow}}+\frac{\Gamma_{L\downarrow}\Gamma_{R\downarrow}D_{\downarrow}}{\Gamma_{L\downarrow}+\Gamma_{R\downarrow}}\right]\left(f_L-f_R\right) 
\end{equation}
and
\begin{equation}
\label{terminal_second_term}
I_{s}^{'}=\int dE\left[\frac{\Gamma_{L\uparrow}}{\Gamma_{L\uparrow}+\Gamma_{R\uparrow}}-\frac{\Gamma_{L\downarrow}}{\Gamma_{L\downarrow}+\Gamma_{R\downarrow}}\right]\tilde{I}_{s}(E)
\end{equation}	
In Eq.(\ref{terminal_second_term}),$\tilde{I}_s(E)$ is defined as $e(2\pi/\hbar)\left\{\tilde{F}_1(E)-\tilde{F}_2(E)\right\}$, where $\tilde{F}_1(E)$, and $\tilde{F}_2(E)$ are the integrands of the variable `E' (terms inside the `[ ]') in Eq.(\ref{F1_final}) and (\ref{F2_final}) respectively. 
%
%
%
%
%
%
The first term in Eq.(\ref{terminal_current}) is the ballistic current ($I_B$) which is clearly a sum of the currents that flow from left to right in the `up-up' and the `down-down' channels. On the other hand, the second term $I_{s}^{'}$ results from the scattering current,$\tilde{I}_{s}(E)$ flowing between the `up-down' channels and gets distributed according to the conductance ratios of left and right.

\begin{figure}[t]
	\centering
	\includegraphics[width=8cm, height=6cm]{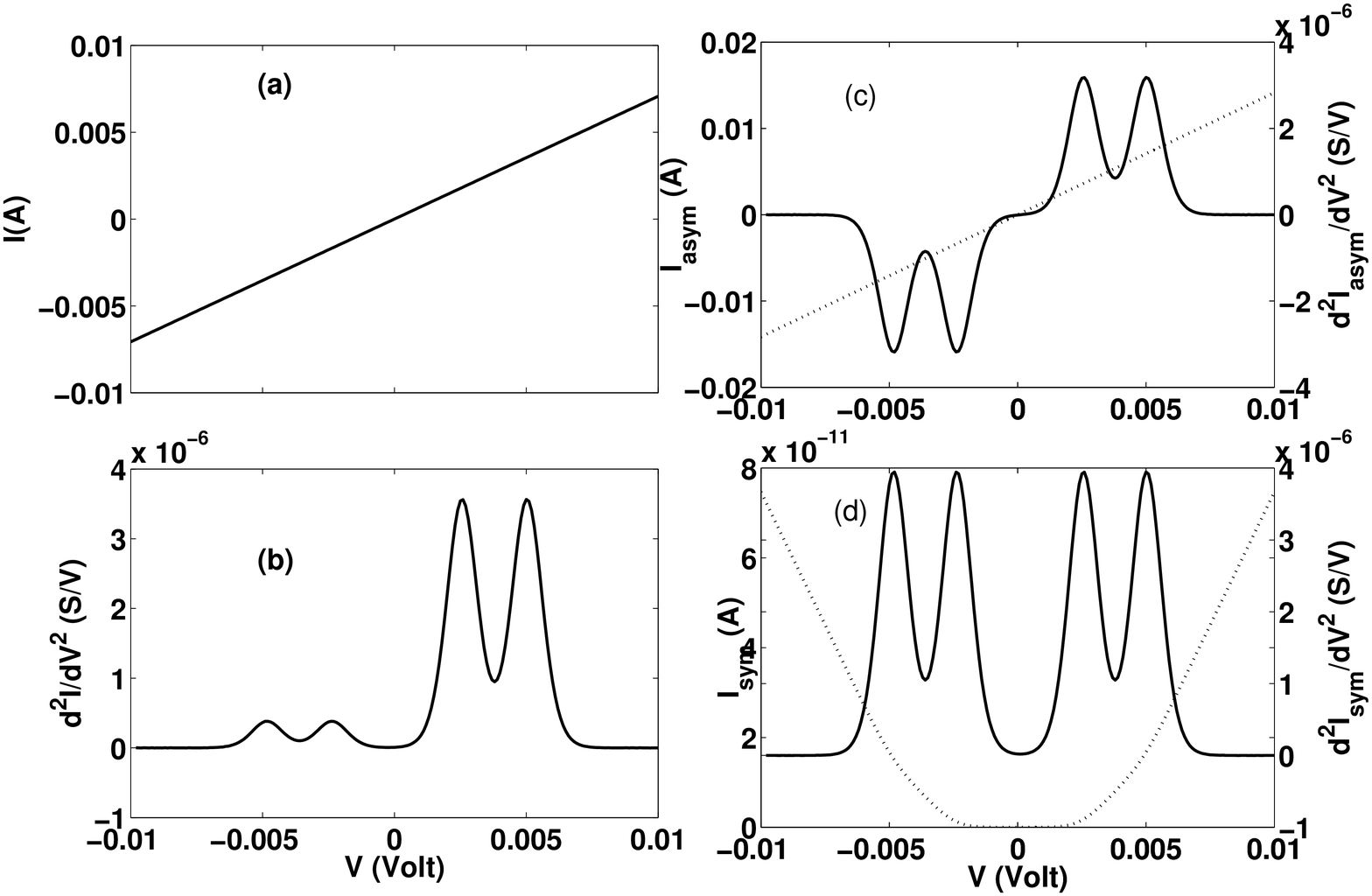}
	\caption{I-V calculated for \emph{Non-Ideal} contact from Eq.(\ref{terminal_current}). The injection efficiencies of the source and drain were assumed to be 60\% and 40\% respectively. Two spin excitations of $\hbar\omega=$ 2.5 and 5 meV were assumed. (a)Total current. No features can be detected due to large contribution from ballistic current. (b) The second derivative of the total current. The peaks retain the information of spin scattering. (c) dotted line---anti-symmetric part of the total current: $I_\text{asym}=I(V)-I(-V)$; solid line---second derivative of the anti-symmetric part. (d) dotted line---symmetric part of the total current: $I_\text{sym}=I(V)+I(-V)$; solid line---second derivative of the symmetric part.  The contact couplings are roughly around 0.1 eV. All other transport parameters are the same as were indicated in Fig.\ref{FIG2}. The temperature was set to T=3K. This is necessary to distinguish among the multiple peaks which will be otherwise washed away because of temperature broadening (see Eq.(\ref{linewidth})).}
	\label{FIG3}
\end{figure}

\emph{$d^2I/dV^2$ as a signature of spin excitation}.-Under small voltage bias (this is the case for all of our treatments), $I_B$ is linear in applied bias,$V$, whereas $I_s^{'}$, having terms containing product of two fermi functions, is quadratic in $V$. Therefore, even though all features coming from spin scattering in the current itself is essentially hidden by a more dominant ballistic part(see Fig.(\ref{FIG3}(a)), the second derivative very effectively retains the signature of the spin excitation (see Fig. \ref{FIG3}(b)). 
For simplicity, we assume the rate of spin flip scattering, which is proportional to $J^2A_{\mp}(\pm\omega)$, is small compared to $\Gamma_{L/R;\uparrow/\downarrow}$, so that $f_\uparrow\approx a_1f_L+b_1f_R$ and $f_{\downarrow}\approx a_2f_L+b_2f_R$, where
\begin{eqnarray}\label{param_values}
a_1=\frac{\Gamma_{L\uparrow}}{\Gamma_{L\uparrow}+\Gamma_{R\uparrow}}\text{; }
b_1=\frac{\Gamma_{R\uparrow}}{\Gamma_{L\uparrow}+\Gamma_{R\uparrow}}\\ \nonumber
a_2=\frac{\Gamma_{L\downarrow}}{\Gamma_{L\downarrow}+\Gamma_{R\downarrow}}\text{; } b_2=\frac{\Gamma_{R\downarrow}}{\Gamma_{L\downarrow}+\Gamma_{R\downarrow}}\nonumber
\end{eqnarray}
We also assume that the density of states is constant in the energy range of interest, which should normally be true for low bias. Taking the second derivative of the terminal current (Eq.(\ref{terminal_current})), noting that the ballistic part is purely linear and recognizing that the voltage enters $F_1$ and $F_2$ only through the fermi functions, we obtain

\begin{equation}\label{secondderivative_general}
\begin{split}
\frac{\partial^2 I}{\partial V^2}&=e^3\left(\frac{2\pi}{\hbar}\right)D^2J_0^2\int d(\hbar\omega)\left[A_{+}(-\omega)-A_{-}(\omega)\right]\\
&\left\{a_1b_2\delta(\hbar\omega+eV)+a_2b_1\delta(eV-\hbar\omega)\right\}
\end{split}
\end{equation}
at $T=0$. Note that (From Eq.(\ref{As})) 
\begin{equation}\label{main_result}
\begin{split}
\left[A_{+}(-\omega)-A_{-}(\omega)\right]&=\sum_{R_j}F(R_j)\sum_{\alpha,\beta}|<\beta|S_-|\alpha>|^2\\
&\delta(E_\beta-E_\alpha-\hbar\omega)(P_\beta-P_\alpha)
\end{split}
\end{equation}
is a measure of the spin excitation spectrum and is reflected in the amplitudes of the two delta functions at $eV=\pm\hbar\omega$, only one of which is non-zero for ideal contacts. This is the main result of this paper. When $T\neq0$ the $\delta$ is broadened into a convolution of two sech$^2$ functions:
\begin{equation}\label{linewidth}
\delta(\hbar\omega-eV)\Rightarrow \int dE \text{ sech}^2\left(\frac{E}{2kT}\right) \text{ sech}^2\left(\frac{\hbar\omega-eV-E}{2kT}\right),
\end{equation}
which describes the numerically computed line shapes in Fig.\ref{FIG3} quite well.

\emph{Symmetrization}.- In the above derivation, we have dropped $\partial^2I_B/\partial V^2$ arguing that $I_B$ vs. V represents the current in the absence of spin-flip scattering and is exactly linear (since $D$ is assumed to be constant). But, if scattering processes that do not flip spin (like inelastic phonon scattering) are present, $\partial^2I_B/\partial V^2$ is not zero. However, it is antisymmetric in V ({See Refs. \cite{wang:ref,galperin-iets:ref}) and therefore, can be eliminated if we take the symmetric component of the second derivative by writing ${\partial^2I_{\text{sym}}}/{\partial V^2}={\partial^2}/{\partial V^2}\left\{I(V)+I(-V)\right\}$, so that the contribution from $\partial^2I_B/\partial V^2$ cancels out.

\emph{Discussion.-} Note that the current calculated in Fig. 2 arises from $N_I=30$ spins in the array interacting with the channel. This current is many orders of magnitude larger than the measurement limit, even though we have used a very conservative estimate for the interaction parameter $J_0$. This means that even single spins should be detectable in realistic structures. 

We believe that a significant strength of the proposed techniques stems from its potential spatial resolution. Since the interaction constant J is determined by the wavefunction overlap of the spins with the channel electrons, we expect the spatial resolution of our proposed method to be limited only by the spread of the wavefunctions of the individual spins, which is the ultimate limit on any measurement technique. For shallow donors the spin wavefunction extend over distances $\approx a_B$ (Bohr radius) which is 100$A^0$ in GaAs.  Deep level traps have even more localized wavefunctions.

One requirement for our proposed device is that the channel length should be much less than the spin-relaxation length due to other causes like spin-orbit coupling. However, this poses no serious restrictions. For example, in doped GaAs, spin relaxation lengths over 1 $\mu$m has been reported \cite{kikka:ref}. Other channel materials of interest like Si, carbon nanotubes etc. are known to have much weaker spin-orbit coupling. 

Ideal half metallic contacts in the AP configuration leads to  $\Gamma_{L\uparrow}=\Gamma_{R\downarrow}\neq0$  and $\Gamma_{L\downarrow}=\Gamma_{R\uparrow}=0$. As seen from Eq.(\ref{terminal_current}), this situation maximizes the effect of $\tilde{I}_s$ on the terminal current ($I_s^{'}=\int dE \tilde{I}_s(E);I_B=0$). If, on the contrary, $\Gamma_{L\uparrow}/\Gamma_{L\downarrow}=\Gamma_{R\uparrow}/\Gamma_{R\downarrow}$, $\tilde{I}_s$ makes no contribution to the terminal current ($I_s^{'}=0$). This corresponds to symmetric unpolarized contacts, which completely destroys the effects we are seeking to exploit. This is in qualitative agreement with a previous result \cite{zhang:ref} that showed, in connection with the `zero bias anomaly' in tunneling magneto resistance (TMR) devices, that spin-flip scattering is more prominent when the spin-polarized contacts are anti-parallel (AP) rather than parallel(P). 

We emphasize the fact that the use of the ferromagnetic contacts make it possible to detect the spin spectra without having to put an external microwave signal. However, the proposed device could be combined with a microwave signal to measure the spin decoherence time $T_2$. Near the resonance condition, the total deviation of the average spin from the direction of static magnetic field depends on the frequency of the rf field, $\omega_{\text{rf}}$ \cite{slichter:ref}. As a result the current as a function of $\omega_{\text{rf}}$ will show a dip whose width is proportional to the transverse decoherence time $T_2$. Note from the example in Fig.2, the changes in the current levels is in the order of nA, which is easily detectable \cite{jiangpaper:ref}. 

Our basic result for the current (see Eq.(\ref{terminal_current})) can be understood by combining a golden rule treatment of the spin-flip current(see Eq.(\ref{scattering_current})) with a heuristic rule for distributing it between the contacts (see Eq.(\ref{terminal_second_term})). Note, however, that this result is obtained rigorously from a NEGF treatment. We have also obtained similar results directly from a Landauer-type scattering formulation with entangled `channel electron'- `spin array' states \footnote{ a longer version showing these comparisons will be published elsewhere.}.

In summary, we have proposed a technique which should allow one to detect spin excitation spectra of a spin array. Compared to existing techniques our proposal has three attractive features: (a) high sensitivity down to single spins, (b) high spatial resolution limited only by the wavefunctions of the individual spins and (c) a relatively common device structure to which a lot of current research is devoted \cite{harris:ref}. Possible applications include single spin detection,  relaxation time measurements, biological sensing of miniature magnetic particles and read out devices for quantum computing, high density memory \cite{gurney:ref}, spin-wave interconnect and logic applications \cite{covington:ref}.

The authors would like to thank D. Sen, A. Ghosh, D. Kienle, P. Srivastava and L.Siddiqui for helpful suggestions. This work was supported by the MARCO focus center for Materials, Structure and Devices.
%
\bibliography{detector_condmat}
\end{document}